\documentclass[12pt,draftclsnofoot,journal,onecolumn]{IEEEtran}

\usepackage{cite}
\usepackage{graphicx}
\usepackage{amsmath}
\usepackage{amsfonts}
\usepackage{array}
\usepackage{amssymb}
\usepackage{subfigure}
\newtheorem{theorem}{Theorem}
\newtheorem{corollary}{Corollary}
\newtheorem{remark}{Remark}

\begin{document}

\title{Multi-mode Transmission for the MIMO Broadcast Channel with Imperfect Channel State Information}
\author{Jun Zhang, Marios Kountouris, Jeffrey G. Andrews, and Robert W. Heath Jr.
\thanks{J. Zhang, J. G. Andrews, and R. W. Heath Jr. are with the Wireless Networking and Communications
Group, Department of Electrical and Computer Engineering, The University of Texas at Austin (Email: jzhang06@mail.utexas.edu, \{jandrews, rheath\}@ece.utexas.edu). M. Kountouris is with SUPELEC, France (Email: marios.kountouris@supelec.fr). This work has been supported in part by AT\&T Labs, Inc. A conference version appeared in \emph{Proc. IEEE Int. Symp. Information Theory 2009}. Last updated: \today.}}

\maketitle

\begin{abstract}
This paper proposes an adaptive multi-mode transmission strategy to improve the spectral efficiency achieved in the multiple-input multiple-output (MIMO) broadcast channel with delayed and quantized channel state information. The adaptive strategy adjusts the number of active users, denoted as the \emph{transmission mode}, to balance transmit array gain, spatial division multiplexing gain, and residual inter-user interference. Accurate closed-form approximations are derived for the achievable rates for different modes, which help identify the active mode that maximizes the average sum throughput for given feedback delay and channel quantization error. The proposed transmission strategy is combined with round-robin scheduling, and is shown to provide throughput gain over single-user MIMO at moderate signal-to-noise ratio. It only requires feedback of instantaneous channel state information from a small number of users. With a feedback load constraint, the proposed algorithm provides performance close to that achieved by opportunistic scheduling with instantaneous feedback from a large number of users.
\end{abstract}

\begin{keywords}
MIMO systems, space division multiplexing, broadcast channels, feedback, delay effects.
\end{keywords}

\section{Introduction}
Multi-user MIMO (MU-MIMO), which allows the base station (BS) to communicate with multiple mobile users simultaneously, provides spatial division multiplexing gain\footnote{In this paper, we use the \emph{spatial division multiplexing gain} to denote the spatial multiplexing gain provided by MU-MIMO to differentiate it from that provided by the point-to-point MIMO.} proportional to the number of antennas ($N_t$) at the BS even with single-antenna mobiles \cite{GesKou07SPMag}, i.e. the sum throughput grows linearly with $N_t$ at high SNR. To exploit the full spatial division multiplexing gain in the MIMO broadcast channel (MIMO-BC), channel state information at the transmitter (CSIT) is required to separate the spatial channels for different users. CSIT, however, is difficult to obtain and is never perfect. Imperfect CSIT causes residual inter-user interference, which degrades the performance of the MIMO-BC \cite{LapSha05Allerton}. The capacity region of the MIMO-BC with imperfect CSIT has not been fully discovered. Although not capacity achieving, linear precoding is a practical transmission technique for the MIMO-BC; it is able to provide full spatial division multiplexing gain given CSIT \cite{YooGol06JSAC,LeeJin07IT}.

\emph{Limited feedback}, in which users feed back quantized channel information through a feedback channel, is an efficient way to provide partial CSIT for the MIMO-BC \cite{LovHea08JSAC}. It has been shown that the full spatial division multiplexing gain in the MIMO-BC can be obtained with carefully designed feedback strategies and a feedback rate that grows linearly with SNR (in dB) and the number of transmit antennas \cite{Jin06IT,YooJin07JSAC,CaiJin07Submit,DinLov07Tsp}. Therefore, linear precoding combined with limited feedback provides a feasible way to exploit the spatial division multiplexing gain in the MIMO-BC.

\subsection{Related Work}
In existing commercial wireless communication systems, the number of feedback bits for each user is fixed and cannot be adjusted with SNR. In addition, there are other CSIT imperfections, such as estimation error and feedback delay, all of which make the system throughput-limited at high SNR due to the residual inter-user interference \cite{Jin06IT,DinLov07Tsp}. One approach to improve the CSIT accuracy in limited feedback systems is to employ multiuser diversity, as proposed in \cite{YooJin07JSAC}. To reduce feedback overhead in such systems, threshold based feedback can be adopted \cite{HuaHea07Tsp}. The CSIT accuracy of limited feedback can also be improved with a progressive refinement of precoding vectors by taking advantage of the temporal channel coherence \cite{HeaWu09}. An alternative approach to deal with imperfect CSIT is adaptively switching between the single-user (SU) and multi-user (MU) modes\footnote{Note that the term ``mode'' used in this paper denotes the number of active users rather than different MIMO transmission techniques, as spatial multiplexing/diversity mode \cite{HeaPau05Tcomm}, or different number of data streams for a given user, as in \cite{LovHea05Tsp}.}, as the SU mode does not suffer from residual interference at high SNR. SU/MU mode switching algorithms for the random orthogonal beamforming system were proposed in \cite{YeuPar07Accept,KouGes07SPAWC}, where each user feeds back its preferred mode and the channel quality information (CQI). For the MIMO downlink with the number of receive antennas greater than or equal to the number of transmit antennas, an adaptive SU/MU mode switching algorithm was proposed in \cite{LeeCha07}, which also considered correlation at transmit antennas.

Most prior mode switching algorithms were based on instantaneous CSIT; no explicit \emph{mode switching point} has been obtained that is related to the key system parameters. An alternative approach is to switch based on the channel distribution information. In \cite{DaiLiu08JSAC}, the number of active users for a limited-feedback zero-forcing (ZF) precoding system was optimized through asymptotic analysis to maximize the spectral efficiency, based on average SNR and channel quantization codebook size. In \cite{ZhaAnd}, a SU/MU mode switching algorithm was proposed for the ZF precoding system considering delayed and quantized CSIT. The mode switching point can be explicitly derived based on the parameters including average SNR, normalized Doppler frequency, and codebook size, which are computable at the BS.

Both switching algorithms in \cite{DaiLiu08JSAC} and \cite{ZhaAnd} require only the selected users to feed back their instantaneous channel information, which is desirable for systems with a large number of users. The technique in \cite{ZhaAnd} is based on non-asymptotic analysis, and thus can better characterize the system behavior with different system parameters, e.g. the operating regions for both SU and MU modes can be determined for different delay/mobility or different codebook sizes. However, it only switches between the SU mode and the MU mode serving as many users as transmit antennas. It neglects the impact of the number of active users on transmit array gain, spatial division multiplexing gain, and residual inter-user interference.

\subsection{Contributions}
In this paper, we propose a general multi-mode switching algorithm that adaptively selects the number of active users in a MIMO-BC considering delayed and quantized CSIT. We derive accurate closed-form approximations for the achievable rates for different modes, based on which the mode that provides the highest throughput can be selected. Such multi-mode transmission (MMT) strategy improves the spectral efficiency by balancing transmit array gain, spatial division multiplexing gain, and residual interference.

The proposed MMT strategy can be combined with channel-independent scheduling algorithms, such as round-robin scheduling, to serve a large number of users. In this way, the scheduling is of low complexity, and it greatly reduces the amount of feedback as only the pre-selected users need to feed back their instantaneous channel information for the precoder design. As instantaneous CSIT is not exploited, the proposed algorithm cannot provide multiuser diversity gain, but it is still able to provide a throughput gain over SU-MIMO transmission and can serve multiple users simultaneously. In addition, in certain scenarios such as with a feedback overhead constraint it is able to provide performance close to that achieved by opportunistic scheduling based on instantaneous CSI feedback from a large number of users.

Based on analytical and numerical results, we provide the following insights and guidelines for the design of MIMO transmission in the broadcast channel with imperfect CSIT:
\begin{enumerate}
\item The SU mode should be applied at both low and high SNRs; at medium SNR, the MU mode can be used with the number of active users adaptively adjusted. The full MU mode that serves $N_t$ users normally should not be activated, as it has the highest residual interference and no array gain.
\item For practical systems such as 3GPP LTE, the MU mode should only be used with low mobility and the number of feedback bits should be increased or advanced feedback techniques such as progressive refinement \cite{HeaWu09} should be employed. A short frame length is desirable for MU-MIMO, as it is related to the amount of CSIT delay.
\item Considering delay-mobility and channel quantization, our analysis can determine which source dominates the CSIT imperfection. For given delay-mobility, we should provide a sufficiently large codebook size, as otherwise channel quantization error will dominate the performance; on the other hand, there is no need for a too large codebook size, as other CSIT imperfections such as delay-mobility start to dominate the performance.
\item For a fixed transmission mode, MMSE precoding outperforms ZF precoding. With imperfect CSIT, both precoding schemes require multi-mode transmission to improve the achievable rate, and they provide close performance. As ZF precoding with MMT requires less feedback and its preferred mode can be easily calculated, it is preferred to MMSE precoding.
\item If there is a constraint on the total number of feedback bits, high-quality feedback from less than $N_t$ users together with mode selection provides good performance compared to instantaneous feedback from a large number of users.
\end{enumerate}

Compared to our previous study in \cite{ZhaAnd}, this paper has made important extensions. First, we have explicitly considered the impact of the number of active users on the performance of MU-MIMO, while \cite{ZhaAnd} only considered the full MU mode with $N_t$ users. Second, the proposed MMT strategy can be easily extended to serve a large number of users and provide a significant throughput gain, while the dual-mode switching in \cite{ZhaAnd} is inflexible and is confined in a system with $N_t$ users. Third, the analysis and simulation results in this paper provide useful insights for the practical system design, while \cite{ZhaAnd} emphasized analysis.

\paragraph*{Organization} The rest of the paper is organized as follows: the system model and
some assumptions are presented in Section \ref{Sec:SysMod}. In
Section \ref{Sec:MultiMode}, closed-form approximations are derived
for the average achievable rates for different modes. User
scheduling based on MMT is proposed in Section
\ref{Sec:Schedule}. Numerical results and conclusions are in Section
\ref{Sec:Num} and \ref{Sec:Conclusion}, respectively.

\paragraph*{Notation} In this paper, we use uppercase boldface letters for matrices ($\mathbf{X}$) and lowercase boldface for vectors ($\mathbf{x}$). $\mathbb{E}[\cdot]$ is the expectation operator. The conjugate transpose of a matrix $\mathbf{X}$ (vector $\mathbf{x}$) is $\mathbf{X}^*$ ($\mathbf{x}^*$). Similarly, $\mathbf{X}^\dag$ denotes the pseudo-inverse, $\tilde{\mathbf{x}}$ denotes the normalized vector of $\mathbf{x}$, i.e. $\tilde{\mathbf{x}}=\frac{\mathbf{x}}{\|\mathbf{x}\|}$, and $\hat{\mathbf{x}}$ denotes the quantized vector of
$\tilde{\mathbf{x}}$.

\section{System Model}\label{Sec:SysMod}
We consider a MIMO-BC with $N_t$ antennas at the
transmitter and $U$ single-antenna mobiles. The discrete-time
complex baseband received signal at the $u$-th user at time $n$ is
given as
\begin{equation}
{y}_u[n]=\mathbf{h}^*_u[n]\sum_{v=1}^{M}\mathbf{f}_{v}[n]{x}_{v}[n]+{z}_u[n],
\end{equation}
where $M$ is the active mode, i.e. the number of active users,
$M=1,2,\cdots,N_t$, $\mathbf{h}_u[n]$ is the channel vector for the
$u$-th user, $\mathbf{h}_u[n]\in\mathbb{C}^{(N_t\times{1})}$, and
${z}_u[n]$ is normalized complex additive Gaussian noise,
$z_u[n]\sim\mathcal{CN}(0,1)$. ${x}_u[n]$ and $\mathbf{f}_u[n]$ are
the transmit signal and precoding vector for the $u$-th user, and
$\mathbf{f}_u[n]\in\mathbb{C}^{(N_t\times{1})}$. The transmit power
constraint is $\mathbb{E}\left[\sum_{u=1}^M|x_u[n]|^2\right]=P$, and
we assume equal power allocation among different users. As the noise
is normalized, $P$ is also the average SNR. Eigen-beamforming is
applied for the SU mode ($M=1$), which transmits along the channel
direction and is optimal for SU-MIMO with perfect CSIT
\cite{Telatar99}. ZF precoding is used for the MU mode
($1<M\leq{N_t}$), as it is possible to derive closed-form results
due to its simple structure, and it is optimal among the set of all
linear precoders at asymptotically high SNR \cite{LeeJin07IT}.

To assist the analysis, we assume that the channel $\mathbf{h}_u[n]$
is well modeled as a spatially white Gaussian channel, with entries
${h}_{i}[n]\sim\mathcal{CN}(0,1)$. The investigation of other
channel models is left to future work. We assume perfect channel
state information (CSI) at the receiver, while the transmitter
obtains CSI through limited feedback from the receiver. In addition,
there is delay in the available CSIT. The models of the CSI delay
and limited feedback are presented as follows.

\subsection{CSI Delay Model}
We consider a stationary ergodic Gauss-Markov block fading regular
process (or auto regressive model of order $1$) \cite[Sec.
16--1]{Hay96}, where the channel remains constant for a symbol
duration and changes from symbol to symbol according to
\begin{equation}\label{ChDelay}
\mathbf{h}[n]=\rho{\mathbf{h}}[n-1]+\mathbf{e}[n],
\end{equation}
where $\mathbf{e}[n]$ is the channel error vector, with independent and identically-distributed (i.i.d.) entries
$e_{i}[n]\sim\mathcal{CN}(0,\epsilon_e^2)$, uncorrelated with $\mathbf{h}[n-1]$, and i.i.d. in time. We assume the CSI delay is one symbol period, but the results can be easily extended to the case with delays of multiple symbols. For the numerical analysis, we use a Gauss-Markov ``fit'' of the classical Clarke's isotropic scattering model \cite{Cla68}, as the true Clarke's model is complicated for analysis. Then the correlation coefficient is $\rho=J_0(2\pi{f_d}T_s)$ with Doppler spread $f_d$, where $T_s$ is the symbol duration and $J_0(\cdot)$ is the zero-th order Bessel function of the first kind. The variance of the error vector is $\epsilon_e^2=1-\rho^2$. The value $f_dT_s$ is the normalized Doppler frequency. As shown in \cite{CaiJin07Submit}, the behaviors are different for the Clarke's and Gauss-Markov model, especially at high SNR. However, the impact of channel quantization still persists and dominates at high SNR, so the conclusion in the paper still holds. In addition, the analysis for the Gauss-Markov model can be extended for the estimation or prediction error, which makes our results more general.

\subsection{Channel Quantization}\label{Sec_Q}
The channel direction information is fed back using a quantization
codebook known at both the transmitter and receiver. The
quantization is chosen from a codebook of unit norm vectors of size
$L=2^B$. Each user is assumed to have a different codebook to avoid
users sharing the same quantization vector\footnote{A low-complexity
method to generate different codebooks for different users is to
first generate a common codebook, and then the codebook for each
user is generated through random unitary rotation of this common
codebook}. The codebook for user $u$ is
$\mathcal{C}_u=\{\mathbf{c}_{u,1},\mathbf{c}_{u,2},\cdots,\mathbf{c}_{u,L}\}$.
Each user quantizes its channel direction to the closest codeword,
measured by the inner product. Therefore, the index of channel for
user $u$ is
\begin{equation}
I_u=\arg\max_{1\leq{\ell}\leq{L}}|\tilde{\mathbf{h}}_u^*\mathbf{c}_{u,\ell}|,
\end{equation}
where $\tilde{\mathbf{h}}_u=\frac{\mathbf{h}_u}{\|\mathbf{h}_u\|}$
is the channel direction. Random vector quantization (RVQ)
\cite{SanHon04ISIT,Jin06IT} is used to facilitate the analysis,
where each quantization vector is independently chosen from the
isotropic distribution on the $N_t$-dimensional unit sphere. We analyze performance averaged
over realizations of all such random codebooks in addition to averaging
over the fading distribution. It was
shown in \cite{DaiLiu08IT} that the RVQ-based codebook is
asymptotically optimal in probability as $N_t$,
$B\rightarrow\infty$, with $\frac{B}{N_t}$ keeping a constant.

Let
$\cos\theta_u=|\tilde{\mathbf{h}}_u^*\hat{\mathbf{h}}_u|$, where $\theta_u=\angle\left(\tilde{\mathbf{h}}_u,\hat{\mathbf{h}}_u\right)$,
then we have \cite{YueLov07Twc}
\begin{equation}\label{eq:xi}
\xi_u=\mathbb{E}_{\theta_u}\left[\cos^2\theta_u\right]=1-L_{i,j}\cdot\beta\left(2^B,\frac{N_t}{N_t-1}\right),
\end{equation}
where $\beta(x,y)$ is the Beta function, i.e.
$\beta(x,y)=\frac{\Gamma(x)\Gamma(y)}{\gamma(x+y)}$ with
$\Gamma(x)=\int_0^\infty{t}^{x-1}e^{-t}\mbox{d}t$ as the Gamma function.

\section{Throughput Analysis and Mode Selection}\label{Sec:MultiMode}
In this section, we derive average achievable rates for different
transmission modes. It is shown that the number of active users to maximize the sum throughput is
closely related to transmit array gain, spatial division
multiplexing gain, and residual inter-user interference. MMT is proposed to adaptively select the active mode to
balance between these effects and maximize the throughput. For a selected mode $M^\star$, $M^\star$ users will be selected based on channel-independent scheduling algorithms such as round-robin scheduling. The scheduling algorithm will be further discussed in Section \ref{Sec:Schedule}.

\subsection{Perfect CSIT}
We first consider the system with perfect CSIT, which serves as the
basis for the analysis of the impact of imperfect CSIT.

\subsubsection{SU-MIMO (Eigen-beamforming), $M=1$}
With perfect CSIT, the beamforming (BF) vector is the channel
direction, i.e.  $\mathbf{f}[n]=\tilde{\mathbf{h}}[n]$. The average
throughput is the same as that of a maximal ratio combining diversity
system, given in \cite{AloGol99Tvt} as
\begin{align}\label{C_BF}
R_{CSIT}(1)&=R_{BF}(\gamma,N_t)\triangleq\mathbb{E}_{\mathbf{h}}\left[\log_2\left(1+\gamma|\mathbf{h}^*[n]\mathbf{f}[n]|^2\right)\right]\notag\\
&=\log_2(e)e^{1/\gamma}\sum_{k=0}^{N_t-1}\frac{\Gamma(-k,1/\gamma)}{\gamma^k},
\end{align}
where $\Gamma(\alpha,x)=\int_x^\infty{t^{\alpha-1}e^{-t}dt}$ is the
complementary incomplete gamma function, and $R_{BF}(\gamma,n)$ is
the rate function for the diversity system with SNR $\gamma$ and
diversity order $n$. The BF system provides transmit array
gain $N_t$ as $\mathbb{E}_{\mathbf{h}}\left[\gamma|\mathbf{h}^*[n]\mathbf{f}[n]|^2\right]=N_t\gamma$. The array gain is defined as the increase in the average combined SNR over the average SNR on each branch \cite{Gold05}.

\subsubsection{MU-MIMO (Zero-forcing), $1<M\leq{N_t}$}
The received SINR for the $u$-th user in a linear precoding MU-MIMO
system in mode $M$ is given by
\begin{align}
\mbox{SINR}_{u}(M)=\frac{\frac{\gamma}{M}|\mathbf{h}^*_u[n]\mathbf{f}_u[n]|^2}{1+\frac{\gamma}{M}\sum_{v\neq{u}}|\mathbf{h}_u^*[n]\mathbf{f}_{v}[n]|^2}.
\end{align}

Denote
${\mathbf{H}}[n]=[{\mathbf{h}}_1[n],{\mathbf{h}}_2[n],\cdots,{\mathbf{h}}_U[n]]^*$,
and the pseudo-inverse of ${\mathbf{H}}[n]$ as
$\mathbf{F}[n]=\mathbf{H}^\dag[n]=\mathbf{H}^*[n](\mathbf{H}[n]\mathbf{H}^*[n])^{-1}$.
The ZF precoding vector for the $u$-th user is obtained by
normalizing the $u$-th column of $\mathbf{F}[n]$. Therefore,
$\mathbf{h}^*_u[n]\mathbf{f}_{v}[n]=0,\,\forall{u\neq{v}}$, i.e.
there is no inter-user interference and each user gets an equivalent
interference-free channel. The SINR for the $u$-th user becomes
\begin{align}\label{ZFSINR}
\mbox{SINR}_{ZF,u}(M)={\frac{\gamma}{M}|\mathbf{h}^*_u[n]\mathbf{f}_u[n]|^2}.
\end{align}
Due to the isotropic nature of i.i.d. Rayleigh fading, such
orthogonality constraints to precancel inter-user interference
consume $M-1$ degrees of freedom at the transmitter. As a result,
the effective channel gain of each parallel channel is a chi-square
random variable with $2(N_t-M+1)$ degrees of freedom
\cite{LeeJin07IT,JinAnd09Tcomm}, i.e.
$|\mathbf{h}_u^*[n]\mathbf{f}_u[n]|^2\sim\chi^2_{2(N_t-M+1)}$.
Therefore, the channel for each user is equivalent to a diversity
channel with order $(N_t-M+1)$ and effective SNR $\frac{\gamma}{M}$.
The average achievable rate for the $u$-th user in mode $M$ is
\begin{align}\label{R_ZFu}
R_{CSIT,u}(M)=R_{BF}\left(\frac{\gamma}{M},N_t-M+1\right),
\end{align}
The average achievable sum rate for the ZF system of mode $M$ is
\begin{align}\label{R_ZF}
R_{CSIT}(M)&=\sum_{u=1}^MR_{CSIT,u}(M)\notag\\
&\stackrel{(a)}{=}MR_{BF}\left(\frac{\gamma}{M},N_t-M+1\right).
\end{align}
The equality (a) follows the homogeneous nature of the network. When
$M=1$, this reduces to \eqref{C_BF}.

\subsubsection{Mode Selection}
From \eqref{R_ZF}, the system in mode $M$ provides a spatial
division multiplexing gain of $M$ and an array gain of $(N_t-M+1)$
for each user. As $M$ increases, the achievable spatial division
multiplexing gain increases but the array gain decreases. Therefore,
there is a tradeoff between the achievable array gain and the
spatial division multiplexing gain. From \eqref{R_ZF}, the mode that
achieves the highest throughput for the given average SNR can be
determined as
\begin{equation}\label{CSIT_MS}
M^\star=\arg\max_{1\leq{M}\leq{N_t}}R_{CSIT}(M).
\end{equation}
Note that this is a very simple optimization problem, as only $N_t$
values need to be computed and compared. This transmission strategy
that adaptively adjusts the number of active users is denoted as
\emph{multi-mode transmission} (MMT).

\subsection{Imperfect CSIT}\label{Sec:ImperfecCSIT}
In this section, we consider imperfect CSIT, including both delay
and channel quantization. As it is difficult to derive the exact
achievable rate for such a system, we provide accurate closed-form
approximations for mode selection. The average achievable rate for
the SU mode $M=1$ in such a system is provided in \cite{ZhaAnd}.

For the MU mode with delay and quantization, the precoding vectors
are designed based on the quantized channel directions with delay,
which achieve $\hat{\mathbf{h}}^*_u[n-1]\mathbf{f}^{(QD)}_{v}[n]=0,\,\forall{u\neq{v}}$. Letters `$Q$' and `$D$' denote parameters of the system with channel quantization (limited feedback) and delay, respectively.
The SINR for the $u$-th user in mode $M$ is given as
\begin{equation}\label{ZFSINR_QD}
\gamma_{ZF,u}^{(QD)}(M)=\frac{\frac{P}{M}|\mathbf{h}^*_u[n]\mathbf{f}^{(QD)}_u[n]|^2}{1+\frac{P}{M}\sum_{v\neq{u}}|\mathbf{h}_u^*[n]\mathbf{f}^{(QD)}_{v}[n]|^2}.
\end{equation}
We assume the mobile users can perfectly estimate the noise and
interference and feed back these information to the transmitter, so
the achievable rate for the $u$-th user is given as
\begin{equation}
R_{QD,u}(M)=\mathbb{E}_\gamma\left[\log_2(1+\gamma_{ZF,u}^{(QD)}(M))\right].
\end{equation}
The same metric is used in \cite{Jin06IT,DinLov07Tsp,ZhaAnd}.

We first analyze the signal term and the interference term in \eqref{ZFSINR_QD}.

\subsubsection{Signal term}
First, we approximate the signal term as
\begin{align}\label{s_DQ}
P_S&=\frac{P}{M}|\mathbf{h}^*_u[n]\mathbf{f}^{(QD)}_u[n]|^2
=\frac{P}{M}|(\rho_u\mathbf{h}_u[n-1]+\mathbf{e}_u[n])^*\mathbf{f}^{(QD)}_u[n]|^2\notag\\
&\stackrel{(a)}{\approx}\frac{P}{M}\left|\rho_u\mathbf{h}^*_u[n-1]\mathbf{f}^{(QD)}_u[n]\right|^2\notag\\
&=\frac{P}{M}\left|\rho_u\|\mathbf{h}_u[n-1]\|\cdot\tilde{\mathbf{h}}^*_u[n-1]\mathbf{f}^{(QD)}_u[n]\right|^2\notag\\
&\stackrel{(b)}{=}\frac{P}{M}\rho_u^2\|\mathbf{h}_u[n-1]\|^2\cdot\left|(\cos\theta_u\hat{\mathbf{h}}_u[n-1]+\sin\theta_u\mathbf{g}_u[n-1])^*\mathbf{f}^{(QD)}_u[n]\right|^2\notag\\
&\stackrel{(c)}{\approx}\frac{P}{M}\xi_u\rho_u^2\|\mathbf{h}_u[n-1]\|^2\cdot\left|\hat{\mathbf{h}}^*_u[n-1]\mathbf{f}^{(QD)}_u[n]\right|^2,
\end{align}
where step (a) removes $\mathbf{e}_u^*[n]\mathbf{f}^{(QD)}_u[n]$,
which is normally very small compared with the remaining term. In step (b), we write $\tilde{\mathbf{h}}_u[n-1]=(\cos\theta_u)\hat{\mathbf{h}}_u[n-1]+(\sin\theta_u)\mathbf{g}_u[n-1]$, where $\theta_u=\angle\left(\tilde{\mathbf{h}}_u[n-1],\hat{\mathbf{h}}_u[n-1]\right)$ and $\mathbf{g}_u[n-1]$ is orthogonal to $\hat{\mathbf{h}}_u[n-1]$. Step (c) approximates the actual channel direction by the quantized version, which is justified for small quantization error, and approximates $\cos\theta_u^2$ by its expectation $\xi_u$, with $\xi_u$ given in \eqref{eq:xi}. As the quantized channel direction $\hat{\mathbf{h}}_u[n-1]$ is independent of each other, and $\mathbf{f}_u^{(QD)}[n]$ is designed to lie in the nullspace of $\hat{\mathbf{h}}_{v}[n-1]$, $\forall{v\neq{u}}$, similar to the case of perfect CSIT, $\|\mathbf{h}_u[n-1]\|\cdot\left|\hat{\mathbf{h}}^*_u[n-1]\mathbf{f}_u^{(QD)}[n]\right|^2\sim\chi^2_{2(N_t-M+1)}$.
So as the perfect CSIT case it also provides an array gain of $N_t-M+1$ for each user.

\subsubsection{Interference Term}
The residual interference term can be approximated as
\begin{align}\label{eq:I_DQ}
&\frac{P}{M}\sum_{v\neq{u}}\left|\mathbf{h}_u^*[n]\mathbf{f}^{(QD)}_{v}[n]\right|^2\notag\\
=&\frac{P}{M}\sum_{v\neq{u}}\left|(\rho_u\mathbf{h}_u[n-1]+\mathbf{e}_u[n])^*\mathbf{f}^{(QD)}_{v}[n]\right|^2\notag\\
\approx&\frac{P}{M}\sum_{v\neq{u}}\rho_u^2\left|\mathbf{h}_u^*[n-1]\mathbf{f}^{(QD)}_{v}[n]\right|^2+\frac{P}{M}\sum_{v\neq{u}}\left|\mathbf{e}_u^*[n]\mathbf{f}^{(QD)}_{v}[n]\right|^2.
\end{align}
The approximation for the denominator comes from removing the terms
with both $\mathbf{e}_u[n]$ and $\mathbf{f}^{(QD)}_{v}[n]$. As shown
in \cite{ZhaAnd}, the interference term due to quantization,
$|\mathbf{h}_u^*[n-1]\mathbf{f}^{(QD)}_{v}[n]|^2$, can be well
approximated as an exponential random variable with mean
$\delta=2^{-\frac{B}{N_t-1}}$. The interference term due to delay,
$|\mathbf{e}_u^*[n]\mathbf{f}^{(QD)}_{v}[n]|^2$, is also an
exponential random variable with mean $\epsilon_{e,u}^2$ as
$\mathbf{e}_u[n]\sim\mathcal{CN}(\mathbf{0},\epsilon_{e,u}^2\mathbf{I})$,
$|\mathbf{f}_{v}[n]|^2=1$.

Based on the analysis and following the results in \cite{Jin06IT,CaiJin07Submit}, we can derive an upper bound for the rate loss for the $u$-th user due to imperfect CSIT, stated in the following theorem.

\begin{theorem}[Rate loss]
The rate loss for the $u$-th user with imperfect CSIT compared to
that with perfect CSIT is upper bounded by
\begin{equation}\label{R_lb}
R_{CSIT,u}-R_{QD,u}\leq\log_2\Delta_u^{(QD)},
\end{equation}
where $\Delta_u^{(QD)}$ is the average noise plus residual
interference, given by
\begin{align}\label{Delta_QD}
\Delta_u^{(QD)}&=\mathbb{E}\left[1+\frac{P}{M}\sum_{v\neq{u}}|\mathbf{h}_u^*[n]\mathbf{f}^{(QD)}_{v}[n]|^2\right]\notag\\
&=1+\left(1-\frac{1}{M}\right)P\left(\rho_u^22^{-\frac{B}{N_t-1}}+\epsilon_{e,u}^2\right).
\end{align}
\end{theorem}

\begin{proof}
The average noise plus interference $\Delta_u^{(QD)}$ can be derived based on the distribution of the interference term and following the approach in \cite{ZhaAnd}, which gives
\begin{align}
&\mathbb{E}\left|\mathbf{h}_u^*[n]\mathbf{f}_v^{(QD)}[n]\right|^2\notag\\
=&\mathbb{E}\left|\rho_u\mathbf{h}_u^*[n-1]\mathbf{f}_v^{(QD)}[n]\right|^2+\mathbb{E}\left|\mathbf{e}_u^*[n]\mathbf{f}_v^{(QD)}[n]\right|^2\notag\\
=&\rho_u^2\cdot2^{-\frac{B}{N_t-1}}+\epsilon_{e,u}^2.\notag
\end{align}
The upper bound is obtained by following the approach in \cite{Jin06IT,CaiJin07Submit}.
\end{proof}

\begin{remark}
From \eqref{Delta_QD}, we see that residual interference/rate loss
depends on delay, codebook size, $N_t$, and $M$. It increases with
delay, and decreases with codebook size. It also increases with $P$,
which makes the system interference-limited at high SNR. With other
parameters fixed, the residual interference increases as $M$
increases, which means it may not be desirable to serve too many
users.
\end{remark}

The bound analysis in \eqref{R_lb} provides helpful insights on the effects of different system parameters on the rate loss, but it is not accurate enough for mode selection. To accurately characterize the achievable rate, we derive the closed-form approximation for the MU mode in the following theorem.
\begin{theorem}[Average achievable throughput]\label{thm_R_DQ}
The average achievable rate for the $u$-th user in the MU system of
mode $M$ ($M>1$) with both delay and channel quantization can be
approximated by
\begin{equation}\label{R_ZFDQ}
R_{QD,u}(M)\approx\log_2(e)\sum_{i=0}^{N_t-L-1}\sum_{j=1}^2\sum_{k=0}^{L-1}\sum_{l=0}^i\frac{a_k^{(j)}(l+k)!}{l!(i-l)!}\alpha^{l+k-i+1}I_1\left(\frac{1}{\alpha},\frac{\alpha}{\delta_j},i,l+k+1\right),
\end{equation}
where $\alpha=\frac{\xi_u\rho_u^2P}{M}$,
$\delta_1=\frac{\rho_u^2P\delta}{M}$,
$\delta_2=\frac{\epsilon_{e,u}^2P}{M}$, $L=M-1$, $a^{(1)}_i$ and
$a^{(2)}_i$ are given in \eqref{eq_a1} and \eqref{eq_a2}, and
$I_1(\cdot,\cdot,\cdot,\cdot)$ is the integral given in
\eqref{eq:I1} in Appendix \ref{pthm_R_DQ}.
\end{theorem}

\begin{proof}
See Appendix \ref{pthm_R_DQ}.
\end{proof}
\begin{remark}
To calculate \eqref{R_ZFDQ} for a given user, we need only
information about its correlation coefficient
($\rho_u^2=1-\epsilon_{e,u}^2$), codebook size ($B$), and average
SNR ($P$). Such information is normally fixed or changes slowly.
Each user can feed back and update its own information, and then the
BS can calculate the achievable rate. Instead, each user may also
calculate the achievable rate and feeds back the preferred mode
index. Note that the calculation and the mode selection is only done
when the parameter changes, such as path loss change due to
mobility.
\end{remark}

As special cases, the average achievable rate for the system with delay or limited feedback is provided as follows.
\begin{corollary}\label{thm_R_D}
The average achievable rate for the $u$-th user in the delayed/limited-feedback
system of mode $M$ ($M>1$) can be approximated by
\begin{equation}\label{R_ZFD}
R_{u}(M)\approx\log_2(e)\sum_{i=0}^{N_t-L-1}\sum_{l=0}^i{{L+l-1}\choose{l}}\frac{\alpha^{L+l-i}}{\beta^L(i-l)!}\cdot{I_1}\left(\frac{1}{\alpha},\frac{\alpha}{\beta},i,L+l\right),
\end{equation}
where $L=M-1$, $\alpha=\frac{\kappa P}{M}$, $\beta=\frac{(1-\kappa)P}{M}$, with
\begin{equation}
\kappa=\left\{\begin{array}{ll}\rho^2&\mbox{with delay only}\\1-\delta&\mbox{with limited feedback only.}\end{array}\right.
\end{equation}
and $I_1(\cdot,\cdot,\cdot,\cdot)$ is given in \eqref{eq:I1} in Appendix \ref{pthm_R_DQ}.
\end{corollary}
\begin{proof}
See Appendix \ref{pf:thm_R_D}.
\end{proof}

In the following, we provide high SNR approximations for MU modes
that can be used to analyze the performance in the
interference-limited region.
\begin{theorem}[High SNR approximation]\label{thm_RateCeil_DQ}
The average achievable rate for the $u$-th user in the system with
both delay and channel quantization in the MU mode $M>1$ at high SNR
is approximated as
\begin{equation}\label{RateCeil_DQ}
R^h_{QD,u}(M)\approx\log_2(e)\sum_{i=0}^{N_t-L-1}\sum_{j=1}^2\sum_{k=0}^{L-1}\frac{a^{(j)}_k\hat{\alpha}^{k+1}(k+i)!}{i!}I_2\left(\frac{\hat{\alpha}}{\hat{\delta}_j},i,k+i+1\right),
\end{equation}
where $\hat{\alpha}=\rho_u^2$, $a^{(1)}_i$ and $a^{(2)}_i$ are given in \eqref{eq_a1} and \eqref{eq_a2} with $\hat{\delta}_1=\rho_u^2\delta$, $\hat{\delta}_2=\epsilon_{e,u}^2$, $L=M-1$, and $I_2(\cdot,\cdot,\cdot)$ is the integral
\begin{equation}\label{eq:I2}
I_2(a,m,n)=\int_0^\infty\frac{x^m}{(x+a)^n(x+1)}dx,
\end{equation}
for which a closed-form expression can be found in \cite[Sec. 3.8]{ZhangDiss}.
\end{theorem}
\begin{proof}
See Appendix \ref{pthm_RateCeil_DQ}.
\end{proof}
The high SNR result for the system with delay or limited feedback is provided in the
following corollary.
\begin{corollary}\label{thm_RateCeil}
The achievable sum rate for the $u$-th user in the delayed/limited feedback system in
MU mode $M>1$ at high SNR is approximated as
\begin{equation}\label{RateCeil_D}
R^h_{D,u}(M)\approx\log_2(e)\sum_{i=0}^{N_t-L-1}{{L+i-1}\choose{i}}\hat{\alpha}^LI_2(\hat{\alpha},i,L+i),
\end{equation}
where $\hat{\alpha}=\frac{\rho_u^2}{\epsilon_{e,u}^2}$ for delayed system, and $\hat{\alpha}=\frac{1-\delta}{\delta}$ for limited feedback system, $L=M-1$, and
$I_2(\cdot,\cdot,\cdot)$ is given in \eqref{eq:I2}.
\end{corollary}
\begin{proof}
Following the steps in Appendix \ref{pthm_RateCeil_DQ} with
$\delta_1=0$.
\end{proof}

Based on \eqref{R_ZFDQ} and the approximation for the beamforming
system in \cite{ZhaAnd}, the active mode that achieves the highest
average throughput in the system with both delay and channel
quantization is selected according to
\begin{equation}\label{eq_mode}
M^\star=\arg\max_{1\leq{M}\leq{N_t}}R_{QD}(M),
\end{equation}
where $R_{QD}(M)=\sum_{u=1}^MR_{QD,u}(M)$, with $R_{QD,u}(M)$ given
in \eqref{R_ZFDQ}.
\begin{remark}
Considering \eqref{Delta_QD}, \eqref{s_DQ}, and \eqref{R_ZFDQ}, the
mode $M$ is now related to residual interference, transmit array
gain, and spatial division multiplexing gain. The idea of MMT is to balance between these effects to maximize the
system throughput. The mode selection is based on fixed system
parameters -- the number of transmit antennas and the codebook size
-- and slow time-varying channel information -- average SNR and
normalized Doppler frequency.
\end{remark}

\begin{figure}
\centering
\includegraphics[width=4.5in]{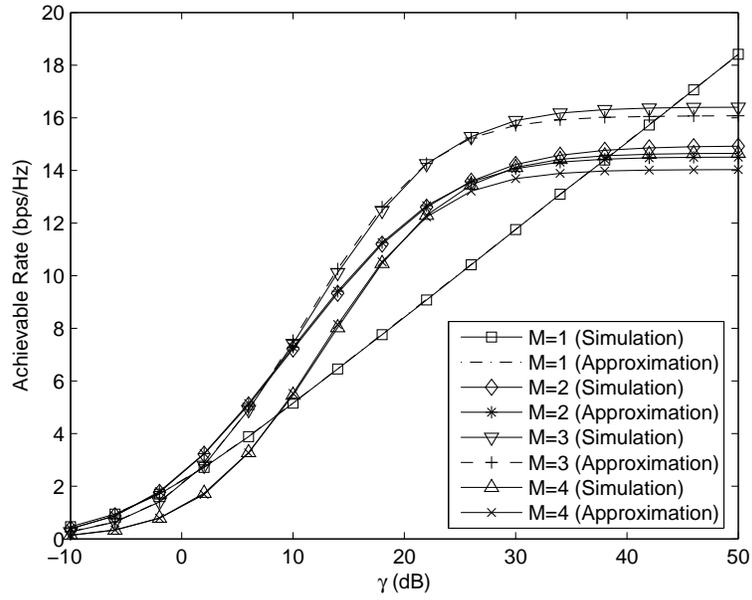}
\caption{Simulation results and approximations for different $M$, $N_t=4$, $v=10$ km/hr, $T_s=1$ msec, $B=18$ bits.}\label{fig_SimvsCal_DQ}
\end{figure}

To show the accuracy of the derived approximations, numerical results are provided in Fig. \ref{fig_SimvsCal_DQ} for different modes, with $N_t=4$, $B=18$ bits, $v=10$ km/hr, and $T_s=1$ msec. We see that the approximation is very accurate at low to medium SNRs. At high SNR, when the sum rate of the MU mode saturates, the approximation becomes a lower bound, and the accuracy decreases as $M$ increases. Interestingly, we see that the mode $M=3$ always provides a
higher throughput than the full MU mode $M=4$. This is due to the fact that the full mode has the highest level of residual interference, as shown in \eqref{Delta_QD}, while it provides no array gain. Therefore, it is desirable to serve fewer than $N_t$ users. We see that MMT is able to provide a throughput gain around $2$ bps/Hz over the dual-mode switching \cite{ZhaAnd} at medium SNR.

\begin{figure}
\centering
\includegraphics[width=4.5in]{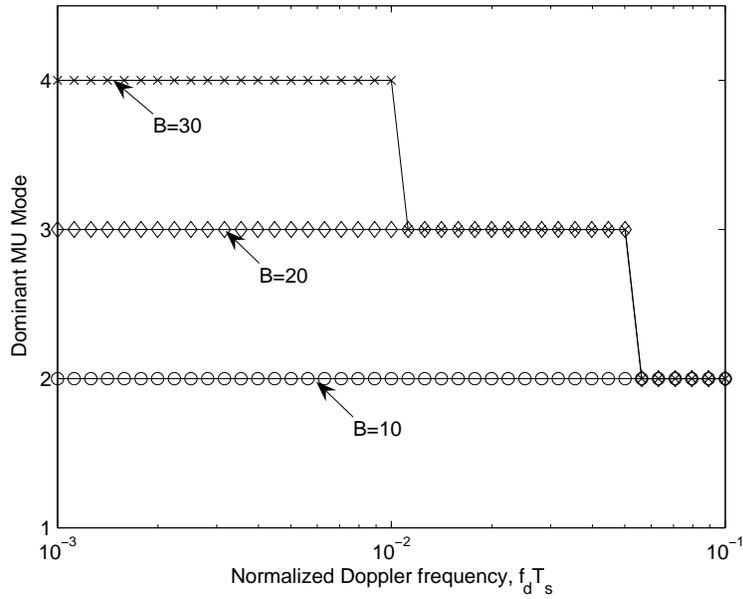}
\caption{The MU mode with the highest rate ceiling for different
$f_dT_s$, with $N_t=4$.}\label{fig_DomMode_DQ}
\end{figure}

It is easy to verify that the high SNR result \eqref{RateCeil_DQ} matches the approximation, and can predict the behavior in the interference-limited region. In Fig. \ref{fig_DomMode_DQ}, we plot $M^*=\arg\max_{M>1}R^h_{QD}(M)$ for different normalized Doppler frequency $f_dT_s$, i.e. the mode with the highest sum rate in the interference-limited region. We see that $M^*$ is different for different $f_dT_s$. For $B=10$ bits, the mode $M=2$ always has the highest
throughput in the considered $f_dT_s$ range, as it provides a higher array gain and has lower residual interference than the higher modes; for $B=20$ bits, as the CSIT accuracy is improved compared to $B=10$ bits, $M=3$ has the highest rate at high SNR when $f_dT_s$ is small, but $M=4$ still has a lower throughput; the highest mode $M=4$ provides the highest throughput only for $B=30$ bits and very small $f_dT_s$ ($<10^{-2}$). So with both delay and channel quantization, the highest mode $M=N_t$ is normally not preferable.

\section{Multi-mode Transmission with Round-robin Scheduling}\label{Sec:Schedule}
For the MIMO-BC with linear precoding, the number of users that can be supported simultaneously is constrained by the number of transmit antennas, so we need to select and transmit to a subset of users in each time slot since typically $U\gg{N_t}$. Based on the throughput analysis in Section \ref{Sec:MultiMode}, MMT can be easily combined with round-robin scheduling, where the scheduling is based on the selected transmission mode but not on the instantaneous CSI feedback. This scheduling algorithm is of low complexity, only requires instantaneous CSI feedback from a small number of users, and provides temporal fairness. It is suitable for delay sensitive services or can be applied when it is possible to get feedback only from few users.

\begin{figure}
\centering
\includegraphics[clip=true,scale=0.8]{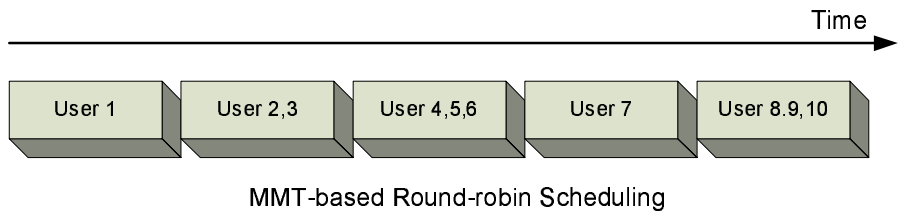}
\caption{Illustration of multi-mode transmission (MMT) based round-robin scheduling.}\label{fig:Illus}
\end{figure}

We assume that the BS has knowledge of the average received SNR and the normalized Doppler frequency of each user, which can be fed back from users and do not require frequent update as they change slowly. Based on such information, the BS can estimate the average achievable throughput for each user for a given subset of selected users based on \eqref{R_ZFDQ}. All the users are indexed and ordered as user $1,2,\ldots,U$. Scheduling starts from the head of the user queue. Once a user is scheduled, it is moved to the tail of the user queue. In the given time slot, denote $u_h$ as the index of the user at the head of the user queue, and $\mathcal{S}$ as the subset of the selected users. The scheduling algorithm is given in Table \ref{Tbl:Scheduling} and also illustrated in Fig. \ref{fig:Illus}. It is similar to the conventional round-robin scheduling and serves users in circular order, but in each time slot multiple users rather than a single user may be served. MMT is applied in each time slot, where the number of selected users is determined based on the analysis in Section \ref{Sec:MultiMode}. Such user scheduling is denoted as \emph{Multi-Mode Transmission (MMT) based scheduling}.

\begin{remark}
Note that for the homogeneous network where all the users have the same average SNR, normalized Doppler frequency, and number of feedback bits, the number of scheduled users in each time slot is fixed, denoted as $M^\star$. The value of $M^\star$ is determined from \eqref{eq_mode}, and then $M^\star$ users are selected from the head of the user queue. This is the main scenario considered in the paper, but the scheduling algorithm can be applied to heterogeneous networks as well.
\end{remark}

Compared to opportunistic scheduling based on instantaneous CSI feedback, the proposed algorithm greatly reduces the amount of CSI feedback, as in each time slot only the selected $M$ users ($M\ll{U}$) need to feed back their instantaneous CSI for the precoder design. In addition, as will be shown in numerical results, it provides performance close to the one with instantaneous CSI feedback from a large number of users. The proposed algorithm is suitable for other systems with scheduling independent of the channel status, such as random selection or the ones based on the queue length.

\section{Numerical Results}\label{Sec:Num}
This section presents numerical results to demonstrate the performance of our proposed transmission strategy and to provide design guidelines in practical systems. We focus on a homogeneous network where all the users have i.i.d. channels. The number of transmit antennas at the BS is $N_t=4$, which is the value currently implemented in broadband wireless standards such as 3GPP LTE.

\subsection{Operating Region and Throughput Gain of Multi-mode Transmission}
In this section, we consider parameters used in the 3GPP LTE (Long Term Evolution) standard \cite{TR36913,TS36211}. The Advanced Wireless Services (AWS) spectrum, which is one of the prime candidates for initial LTE deployment in the US, is considered, i.e. the carrier frequency is $f_c=2.1$ GHz. In LTE, the minimum size of radio resource that can be allocated in the time domain is one subframe of 1 msec. Considering the propagation and processing time, the typical CSIT delay in the FDD mode is five subframes, i.e. the delay is $\tau=5$ msec.

\subsubsection{Operating Regions}
\begin{figure*}
\centering{\subfigure[Different $v$, $f_c=2.1$ GHz, $\tau=5$ msec, $B=15$ bits.]{\includegraphics[width=3in]{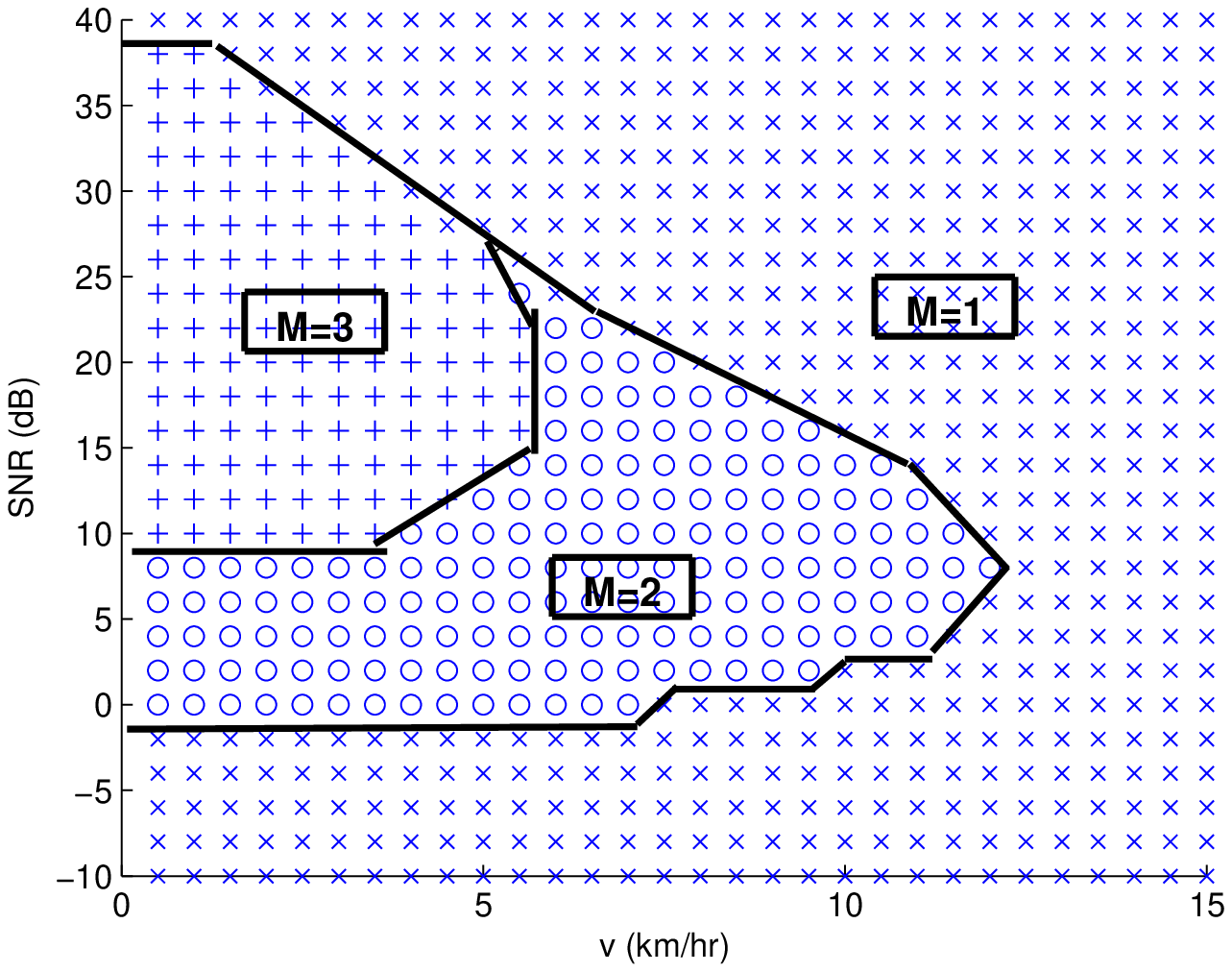} \label{fig:SwitchRegion_D_B18}} \hfil
\subfigure[Different $B$, $f_c=2.1$ GHz, $\tau=5$ msec., and $v=5$ km/hr.]{\includegraphics[width=3in]{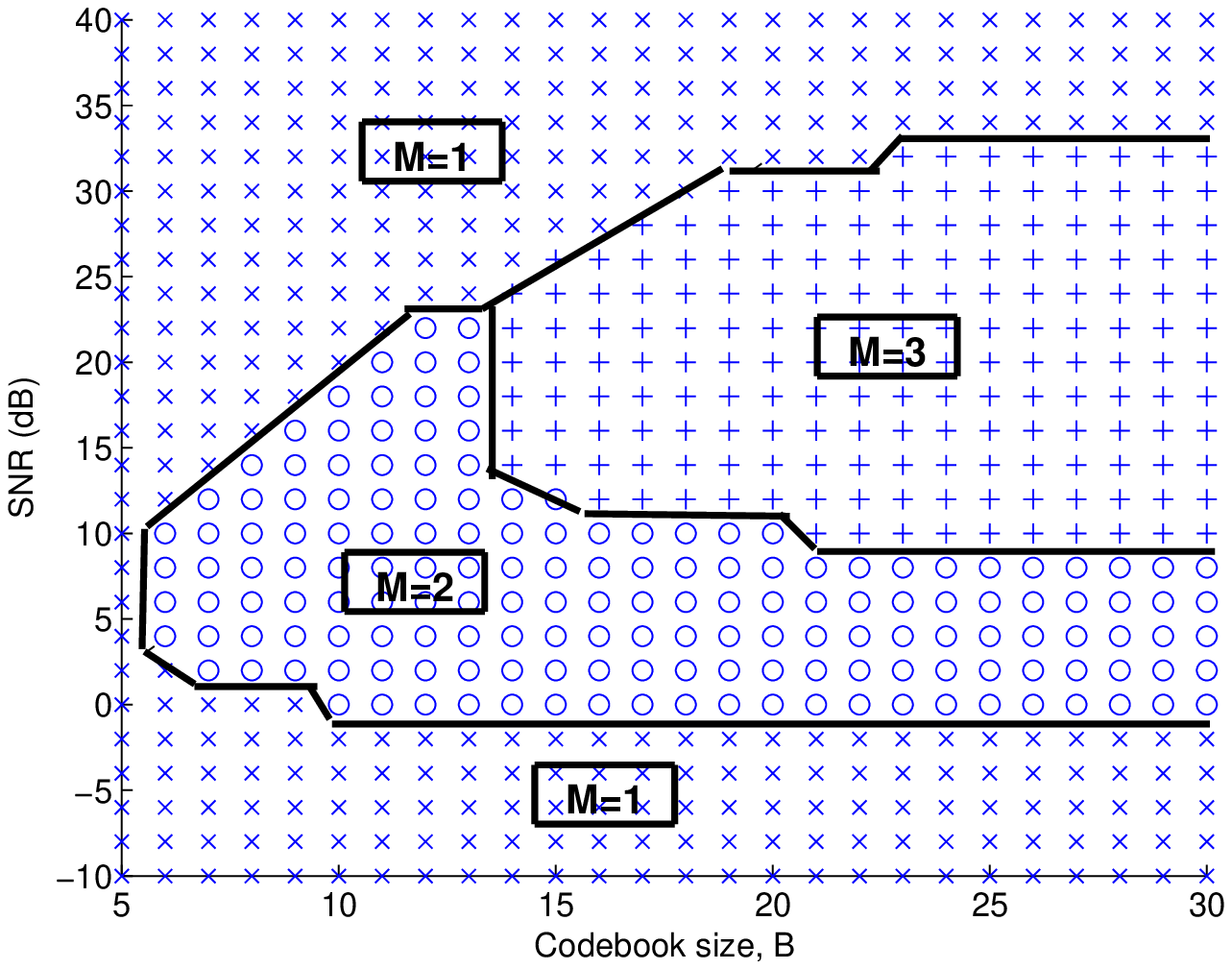} \label{fig:SwitchRegion_Q_v10}}}
\caption{Operating regions for different modes with both CSI delay and channel quantization, $N_t=4$. The mode $M=i$ means that there are $i$ active users. In this plot, `$\times$' is for $M=1$, `o' is for $M=2$, `$+$' is for $M=3$, and `$\Box$' is for $M=4$. Note that the highest mode $M=4$ is never activated in both figures.} \label{fig:MultiMode:SwitchRegion_DQ}
\end{figure*}

Based on the results in Section \ref{Sec:MultiMode}, the preferred mode $M^\star$ can be determined for a given scenario. Accordingly, the operating regions for different modes can be plotted for different system parameters. Fig. \ref{fig:SwitchRegion_D_B18} and Fig. \ref{fig:SwitchRegion_Q_v10} show the operating regions for the system with both delay and channel quantization, for different mobility $v$ and different feedback bits $B$, respectively, where each mark on the figure denotes the type of the active mode. There are several key observations:
\begin{enumerate}
\item For the given $v$ and $B$, the SU mode ($M=1$) will be active at both low and high SNRs, due to its array gain and the robustness to imperfect CSIT, respectively.
\item For MU modes to be active, $v$ needs to be small while $B$ needs to be large. Specifically, to activate $M=2$, we need $v\leq12$ km/hr with $B=15$ bits as in Fig. \ref{fig:SwitchRegion_D_B18}, and need $B\geq6$ bits with $v=5$ km/hr as in Fig. \ref{fig:SwitchRegion_Q_v10}; to activate $M=3$, we need $v\leq6$ km/hr with $B=15$ bits and $B\geq14$ bits with $v=5$ km/hr. Note that in LTE each user only feeds back 4 bits to indicate its channel direction.
\item The full MU mode $M=N_t$ is not active at all with the considered parameters, as it suffers from the highest residual interference and does not provide array gain.
\end{enumerate}

\subsubsection{SU-MIMO vs. MMT}
In Fig. \ref{fig:LTE}, we compare MMT with ZF precoding (MMT-ZF) and the single-user beamforming (SU-BF) transmission, both with round-robin scheduling, and with $f_c=2.1$ GHz, $\tau=5$ msec., and $v=5$ km/hr. We see that for $B=4$ the curves of MMT-ZF and SU-BF overlap, which means no MU mode is activated. This confirms the result in Fig. \ref{fig:SwitchRegion_Q_v10}. For $B=8$, MMT-ZF provides throughput gain over SU-BF for SNR in $0\sim18$ dB. For $B=12$, MMT-ZF provides throughput gain for SNR in $-5\sim25$ dB, which is larger than $20\%$ for SNR=$10\sim20$ dB.

\subsubsection{Delay vs. Quantization Error}
\begin{figure}
\centering
\includegraphics[width=4.5in]{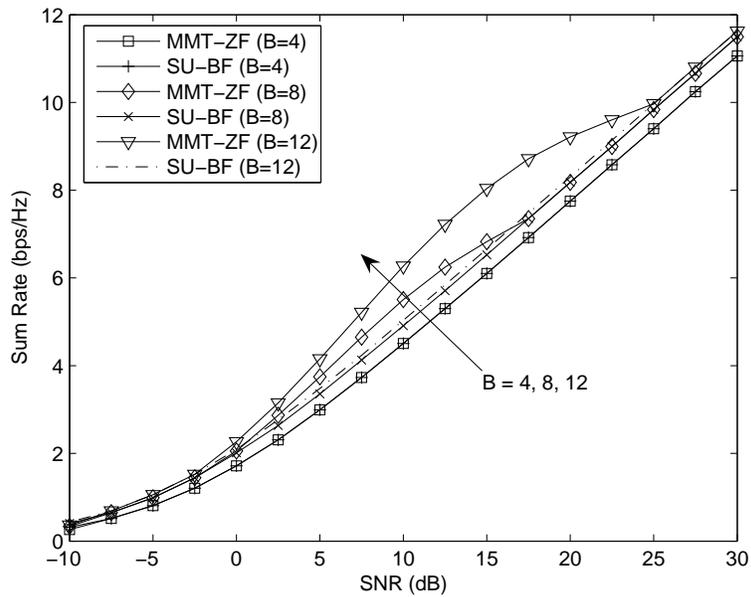}
\caption{Simulation results of SU-BF and MMT-ZF with different $B$, $N_t=4$, $f_c=2.1$ GHz, $\tau=5$ msec., and $v=5$ km/hr.}\label{fig:LTE}
\end{figure}

\begin{figure}
\centering
\includegraphics[width=4.5in]{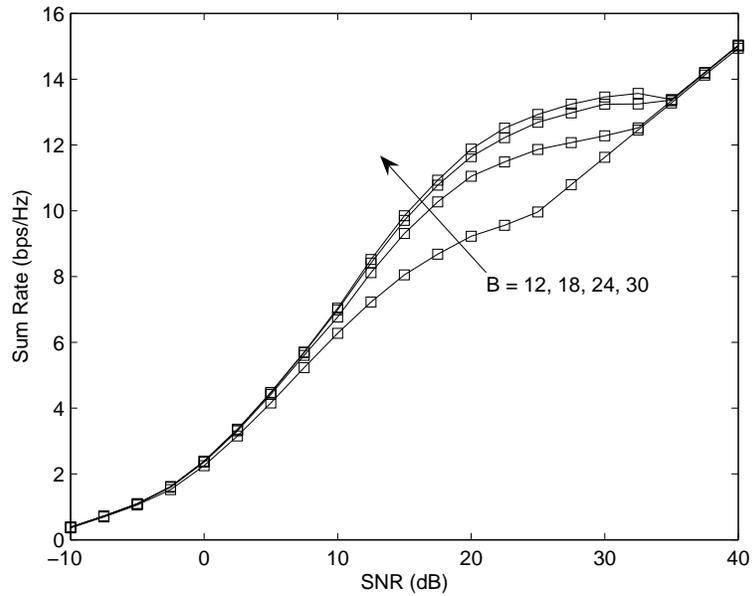}
\caption{Performance of MMT-ZF with different $B$, $N_t=4$, $f_c=2.1$ GHz, $\tau=5$ msec., and $v=5$ km/hr.}\label{fig:DiffB}
\end{figure}

From the analysis of residual interference terms in Section \ref{Sec:ImperfecCSIT}, we can determine for a given scenario which effect dominates, delay-mobility or channel quantization error. Each interference term due to delay-mobility has variance $\sigma_D^2=\epsilon_{e,u}^2$, while each interference term due to channel quantization error has variance $\sigma_Q^2=\rho_u^22^{-\frac{B}{N_t-1}}$. For example, for $v=5$ km/hr, $f_c=2.1$ GHz, and $\tau=5$ msec, we have $\sigma_D^2=0.0458$, and $\sigma_Q^2=0.3787$ with $B=4$, $\sigma_Q^2=0.1503$ with $B=8$, so $\sigma_D^2\ll\sigma_Q^2$ in these scenarios and the channel quantization error dominates the performance, as shown in Fig. \ref{fig:LTE}; to get $\sigma_D^2\approx\sigma_Q^2$, we need $B=13$, but a too large $B$ will not help much as delay-mobility will start to dominate. Fig. \ref{fig:DiffB} shows the performance of ZF-MMT with different values of $B$. We see that the sum rate cannot be further improved once $B$ is sufficiently large ($B>20$).

\begin{remark}
The numerical results in this section provide the following insights:
\begin{enumerate}
\item As shown in Fig. \ref{fig:SwitchRegion_D_B18}, for a given delay and give $B$, the mobility plays a significant role, and MU-MIMO should only be used with low mobility ($\leq12$ km/hr).
\item Both Fig. \ref{fig:SwitchRegion_Q_v10} and Fig. \ref{fig:LTE} show that the number of feedback bits in LTE ($B=4$) is not large enough for MU-MIMO and should be increased ($B\geq8$).
\item As CSI feedback occurs only in certain subframes, the delay in available CSIT is related to the radio frame length. Therefore, it is expected that the MU-MIMO is more applicable in the LTE system which has a shorter frame length (1 msec) than the WiMAX system (5 msec).
\item Reducing quantization error by increasing $B$ is not enough to fully exploit the throughput gain of MU-MIMO, and other CSIT imperfections should be taken into consideration.
\end{enumerate}
\end{remark}

\subsection{ZF vs. MMSE Precoding}
\begin{figure}
\centering
\includegraphics[width=4.5in]{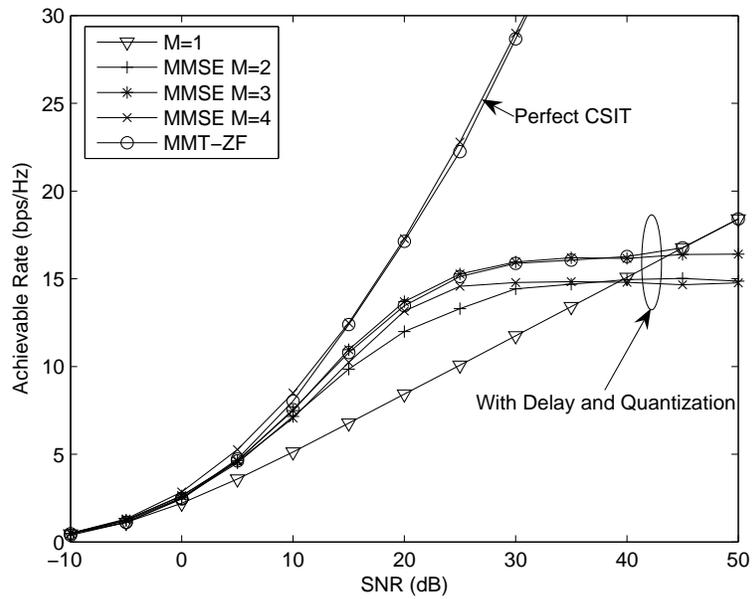}
\caption{Simulation results of MMSE precoding and MMT-ZF systems, $N_t=4$. For imperfect CSIT, $B=18$ bits, $f_c=2.1$
GHz, $\tau=5$ msec., and $v=5$ km/hr.}\label{fig:MMSEvsMMT}
\end{figure}

\emph{MMSE precoding}, or \emph{regularized ZF precoding}, can increase the throughput at low SNR compared to ZF precoding \cite{PeeHoc05Tcomm}. Fig. \ref{fig:MMSEvsMMT} compares the sum rates of MMT-ZF and MMSE precoding. For the \emph{perfect CSIT} case, the number of active users for MMSE precoding is fixed to be $U=N_t$, as little gain can be achieved by varying the user number. We see that the sum rates of the two systems are very close. This means that MMT improves the performance of ZF precoding and approaches that of MMSE precoding. With \emph{imperfect CSIT}, simulation results for different modes with MMSE precoding are plotted, showing that mode switching is also required to improve the spectral efficiency for MMSE precoding. If MMT is applied for both systems, we see that the performance of ZF precoding approaches that of MMSE precoding. Note that MMSE precoding requires instantaneous CSI feedback from all the user, while the number of users that need to feed back instantaneous CSI for ZF precoding depends on the active mode and normally is less than ${N_t}$. In addition, as MMSE precoding is difficult to analyze, the preferred mode cannot be easily determined. Therefore, employing MMT, ZF precoding is preferred to MMSE precoding.

\subsection{MMT Under a Feedback Overload Constraint}
In this section, we consider the scenario with a constraint on the total feedback overhead, i.e. the total number of feedback bits from all the users is fixed to be $B_T$ \cite{RavJin09Submit}. We compare the proposed MMT-ZF with round-robin scheduling and the ZF precoding with opportunistic user selection (US-ZF) that is based on instantaneous CSI feedback. As shown in \cite{RavJin09Submit}, with a feedback overhead constraint it is more desirable to get high-rate/high-quality feedback from a small number of users than low-rate/course feedback from a large number of users, and the optimal number of feedback bits from each user and the number of active users can be determined. We focus on the impact of limited feedback, and delay is not considered in this section. For MMT-ZF, the number of feedback bits for each user is $\lfloor\frac{B_T}{M}\rfloor$ for mode $M$. For US-ZF, the optimal feedback bits for each user, $B^\star_{US}$, is obtained throughput simulation\footnote{In \cite{RavJin09Submit}, an approximation was derived to solve for $B^\star_{US}$ (eq. (11)). However, it is based on the rate loss derivation, and is not accurate especially for medium to high SNR values. The inaccuracy of the rate loss based result at high SNR was also shown in \cite{ZhaAnd}.}, and $\lfloor\frac{B_T}{B^\star_{US}}\rfloor$ users feed back their instantaneous CSI for scheduling.

\begin{figure*}
\centerline{\subfigure[Sum rates of different systems, $T=100$]{\includegraphics[width=3.6in]{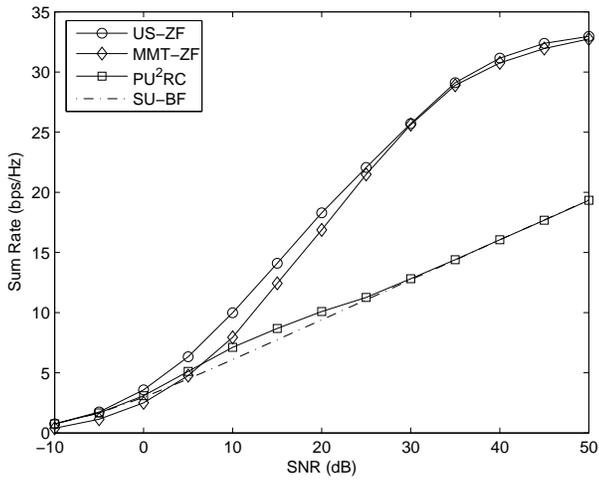} \label{fig:DiffSNRT100}} \hfil
\subfigure[Sum rates for different $B_T$, SNR=$15$ dB.]{\includegraphics[width=3.6in]{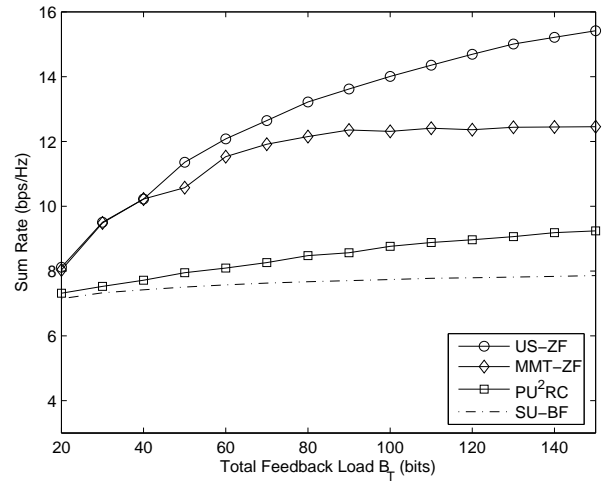} \label{fig:DiffT15dB}}}
\caption{Sum rates of different systems with a feedback overhead constraint. The curve of MMT-ZF is not smooth for different $B_T$ due to the roundoff of $\lfloor\frac{B_T}{M}\rfloor$.} \label{fig:DiffSNR}
\end{figure*}

Fig. \ref{fig:DiffSNR} shows the performance of MMT-ZF and US-ZF, together with the PU$^2$RC (Per Unitary basis stream User and Rate Control) \cite{HuaAnd09Tvt} and single-user beamforming (SU-BF) with channel-dependent maximum rate scheduling. We see that MMT-ZF almost always performs better than both PU$^2$RC and SU-BF, and its performance is close to US-ZF. At low SNR, PU$^2$RC provides slightly higher throughput than MMT-ZF as it provides multiuser diversity, which is a kind of power gain and dominates the performance at low SNR. For a given SNR value, when $B_T$ keeps increasing, the throughput of US-ZF increases due to multiuser diversity, while the sum rate of MMT-ZF saturates as the full mode $M=N_t$ is activated and there is limited performance gain to further improve the accuracy of CSIT.

\section{Conclusions}\label{Sec:Conclusion}
In this paper, we propose a multi-mode transmission strategy that adaptively adjusts the number of active users based on the average achievable rate. Considering transmit array gain, spatial division multiplexing gain and residual inter-user interference, multi-mode transmission improves the spectral efficiency over SU-MIMO at medium SNR. It is shown that the full mode $M^\star=N_t$ will normally not be activated as it has the highest residual interference and no array gain. Multi-mode transmission can be combined with round-robin scheduling to serve a large number of users, which selects users based on average SNR, codebook size, and normalized Doppler frequency. The proposed algorithm significantly reduces the feedback amount, and provides throughput close to the one based on instantaneous CSI feedback when there is a total feedback overhead constraint. The analysis and numerical results provide insights and design guidelines that are of practical importance.

\useRomanappendicesfalse
\appendix

\subsection{Proof of Theorem \ref{thm_R_DQ}}\label{pthm_R_DQ}
Assuming interference terms are independent, and independent of the signal term, $\gamma_{ZF,u}^{(QD)}$ can be approximated as
\begin{equation}\label{SINR_approxDQ}
\gamma_{ZF,u}^{(QD)}\approx\frac{\alpha{z}}{1+\delta_1y_1+\delta_2y_2}\triangleq{x},
\end{equation}
where $\alpha=\frac{\rho_u^2P}{M}$, $\delta_1=\frac{\rho_u^2P\delta}{M}$, $\delta_2=\frac{\epsilon_{e,u}^2P}{M}$, $y_1\sim\chi^2_{2L}$, $y_2\sim\chi^2_{2L}$, $z\sim\chi^2_{2(N_t-L)}$, $L=M-1$, and $y_1$, $y_2$, $z$ are independent of each other.

Let $y=\delta_1y_1+\delta_2y_2$, then the pdf of $y$, which is the sum of two independent chi-square random variables, is given as
\cite{Sim02}
\begin{align}
p_Y(y)=e^{-y/\delta_1}\sum_{i=0}^{L-1}a^{(1)}_iy^i+e^{-y/\delta_2}\sum_{i=0}^{L-1}a^{(2)}_iy^i
=\sum_{j=1}^2\sum_{k=0}^{L-1}a_k^{(j)}y^ke^{-y/\delta_j},
\end{align}
where
\begin{align}
a^{(1)}_i&=\frac{1}{\delta_1^{i+1}(L-1)!}\left(\frac{\delta_1}{\delta_1-\delta_2}\right)^L\frac{(2(L-1)-i)!}{i!(L-1-i)!}\left(\frac{\delta_2}{\delta_2-\delta_1}\right)^{L-1-i}\label{eq_a1},\\
a^{(2)}_i&=\frac{1}{\delta_2^{i+1}(L-1)!}\left(\frac{\delta_2}{\delta_2-\delta_1}\right)^L\frac{(2(L-1)-i)!}{i!(L-1-i)!}\left(\frac{\delta_1}{\delta_1-\delta_2}\right)^{L-1-i}\label{eq_a2}.
\end{align}
We first derive the cumulative distribution function (cdf) of the random variable $x$ as follows.
\begin{align}
F_X(x)&=P\left(\frac{\alpha{z}}{1+y}\leq{x}\right)
=\int_0^\infty{F}_{Z|Y}\left(\frac{x}{\alpha}(1+{y})\right)p_Y(y)dy\notag\\
&=\int_0^\infty\left(1-e^{-\frac{x}{\alpha}(1+{y})}\sum_{i=0}^{N_t-L-1}\frac{\left[\frac{x}{\alpha}(1+{y})\right]^i}{i!}\right)p_Y(y)dy\notag\\
&=1-e^{-x/\alpha}\int_0^\infty\left\{e^{-\frac{x}{\alpha}y}\sum_{i=0}^{N_t-L-1}\frac{\left[\frac{x}{\alpha}(1+y)\right]^i}{i!}\cdot\sum_{j=1}^2\sum_{k=0}^{L-1}a_k^{(j)}y^ke^{-y/\delta_j}\right\}dy\notag\\
&\stackrel{(a)}{=}1-e^{-x/\alpha}\sum_{i=0}^{N_t-L-1}\sum_{j=1}^2\sum_{k=0}^{L-1}\frac{a_k^{(j)}(x/\alpha)^i}{i!}\sum_{l=0}^i{i\choose{l}}\int_0^\infty\exp\left[-\left(\frac{x}{\alpha}+\frac{1}{\delta_j}\right)y\right]y^{l+k}dy\notag\\
&\stackrel{(b)}{=}1-\sum_{i=0}^{N_t-L-1}\sum_{j=1}^2\sum_{k=0}^{L-1}\sum_{l=0}^i\frac{a_k^{(j)}(l+k)!}{l!(i-l)!}\cdot\frac{\alpha^{l+k-i+1}e^{-x/\alpha}x^i}{\left(x+\frac{\alpha}{\delta_j}\right)^{l+k+1}},
\end{align}
where step (a) follows binomial expansion of $(y+1)^i$, and step (b) follows the equality $\int_0^\infty{y}^Me^{-\alpha{y}}=M!\alpha^{-(M+1)}$.

The expectation of $\ln(1+x)$ on $x$ is derived as follows.
\begin{align}\label{eq_AppendC}
\mathbb{E}_X\left[\ln(1+X)\right]&=\int_0^\infty\ln(1+x)dF_X
\stackrel{(a)}{=}\int_0^\infty\frac{1-F_X(x)}{x+1}dx\notag\\
&=\int_0^\infty\left[\sum_{i=0}^{N_t-L-1}\sum_{j=1}^2\sum_{k=0}^{L-1}\sum_{l=0}^i\frac{a_k^{(j)}(l+k)!}{l!(i-l)!}\cdot\frac{\alpha^{l+k-i+1}e^{-x/\alpha}x^i}{\left(x+\frac{\alpha}{\delta_j}\right)^{l+k+1}(x+1)}dx\right]\notag\\
&\stackrel{(b)}{=}\sum_{i=0}^{N_t-L-1}\sum_{j=1}^2\sum_{k=0}^{L-1}\sum_{l=0}^i\frac{a_k^{(j)}(l+k)!}{l!(i-l)!}\alpha^{l+k-i+1}I_1\left(\frac{1}{\alpha},\frac{\alpha}{\delta_j},i,l+k+1\right),
\end{align}
where step (a) follows integration by parts and $I_1(\cdot,\cdot,\cdot,\cdot)$ in step (b) is given by
\begin{equation}\label{eq:I1}
I_1(a,b,m,n)=\int_0^\infty\frac{x^me^{-ax}}{(x+b)^n(x+1)}dx,
\end{equation}
for which a closed-form expression can be found in \cite[Sec. 3.8]{ZhangDiss}.
Then the achievable sum rate for the $u$-th user in mode $M$ can be approximated as
\begin{align}\notag
R_{QD,u}(M)&=\mathbb{E}_\gamma\left[\log_2(1+\gamma_{ZF,u}^{(QD)})\right]
\approx\log_2(e)\cdot\mathbb{E}_X\left[\ln(1+X)\right],
\end{align}
which gives \eqref{R_ZFDQ}.

\subsection{Proof of Corollary \ref{thm_R_D}}\label{pf:thm_R_D}
For the system with delay, similar to \eqref{s_DQ}, the received signal power for the $u$-th user is approximated as
\begin{align}
P_S^{(D)}\approx\frac{P}{M}|\rho_u\mathbf{h}^*_u[n-1]\mathbf{f}_u[n]|^2.
\end{align}
Similar to \eqref{SINR_approxDQ}, the received SINR can be approximated by
\begin{equation}\label{eq:SINR_D}
\gamma_{ZF,u}^{(D)}\approx\frac{\alpha{z}}{1+\beta y}\triangleq{x},
\end{equation}
where $\alpha=\frac{\rho_u^2P}{M}$, $\beta=\frac{\epsilon_{e,u}^2P}{M}$, $y\sim\chi^2_{2L}$, $z\sim\chi^2_{2(N_t-L)}$, $L=M-1$, and $y$, $z$ are independent of each other.

For the system with limited feedback, similar to \eqref{s_DQ}, we get the received signal power as
\begin{align}
P_S^{(Q)}&=\frac{P}{M}|\mathbf{h}^*_u[n]\mathbf{f}^{(Q)}_u[n]|^2\notag\\
&=\frac{P}{M}\|\mathbf{h}_u[n]\|^2\cdot|(\cos\theta_u)\hat{\mathbf{h}}^*_u[n]\mathbf{f}^{(Q)}_u[n]+(\sin\theta_u)\mathbf{g}^*_u[n]\mathbf{f}^{(Q)}_u[n]|^2\notag\\
&\stackrel{(a)}{\approx}\frac{P}{M}\|\mathbf{h}_u[n]\|^2\cdot|(\cos\theta_u)\hat{\mathbf{h}}^*_u[n]\mathbf{f}^{(Q)}_u[n]|^2\notag\\
&\stackrel{(b)}{\approx}\frac{P}{M}\|\mathbf{h}_u[n]\|^2\cdot\mathbb{E}_\theta[\cos\theta_u^2]\cdot|\hat{\mathbf{h}}^*_u[n]\mathbf{f}^{(Q)}_u[n]|^2,
\end{align}
where in step (a) we remove the term with $\sin\theta_u$ which is small for a large $B$, and in step (b) we approximate $\cos\theta_u^2$ by $\mathbb{E}_\theta[\cos\theta^2_u]$. As shown in \cite{Jin06IT}, $\mathbb{E}_\theta[\cos\theta^2_u]$ is well approximated by $1-\delta$, with $\delta=2^{-\frac{B}{N_t-1}}$. Then the received SINR for the $u$-th user is approximated by
\begin{equation}\label{eq:SINR_Q}
\gamma_{ZF,u}^{(Q)}\approx\frac{\alpha{z}}{1+\beta y}\triangleq{x},
\end{equation}
where $\alpha=\frac{(1-\delta)P}{M}$, $\beta=\frac{\delta P}{M}$, $y\sim\chi^2_{2L}$, $z\sim\chi^2_{2(N_t-L)}$, $L=M-1$, and $y$, $z$ are independent of each other.

From \eqref{eq:SINR_D} and \eqref{eq:SINR_Q}, we can get the result in \eqref{R_ZFD} following the steps in Appendix \ref{pthm_R_DQ}.

\subsection{Proof of Theorem
\ref{thm_RateCeil_DQ}}\label{pthm_RateCeil_DQ} From
\eqref{SINR_approxDQ}, the received SINR for the $u$-th user at high
SNR is approximated by
\begin{equation}\notag
\gamma_{ZF,u}^{(D)}\approx\frac{\hat{\alpha}{z}}{\hat{\delta}_1{y_1}+\hat{\delta}_2y_2},
\end{equation}
where $\hat{\alpha}=\rho_u^2$, $\hat{\delta}_1=\rho_u^2\delta$,
$\hat{\delta}_2=\epsilon_{e,u}^2$, $z\sim\chi^2_{2(N_t-L)}$,
$y_1\sim\chi^2_{2L}$, $y_1\sim\chi^2_{2L}$, $L=M-1$, and $z$, $y_1$
and $y_2$ are independent.

Denote
$x=\frac{\hat{\alpha}{z}}{\hat{\delta}_1{y_1}+\hat{\delta}_2y_2}$,
following the same steps as in Appendix \ref{pthm_R_DQ}, the cdf of
$x$ is derived as
\begin{equation}
F_X(x)=1-\sum_{i=0}^{N_t-L-1}\sum_{j=1}^2\sum_{k=0}^{L-1}\frac{a_k^{(j)}\hat{\alpha}^{k+1}(k+i)!}{i!}\cdot\frac{x^i}{\left(x+\frac{\hat{\alpha}}{\hat{\delta}_j}\right)^{k+i+1}}.
\end{equation}
Then following the steps in \eqref{eq_AppendC} we get the result in
\eqref{RateCeil_DQ}.

\bibliographystyle{IEEEtran}
\bibliography{CSIDelay}

\begin{table}
\caption{Scheduling Algorithm}\label{Tbl:Scheduling} \centering
\begin{enumerate}
\item Initially, set $\mathcal{S}=\{u_h\}$, $R_{old}=R_{QD}(u_h)$, $\hat{u}=u_h$.
\item While $|\hat{u}-u_h|<N_t$
\begin{enumerate}
\item $\hat{\mathcal{S}}=\hat{\mathcal{S}}+\{\hat{u}+1\mbox{ mod }U\}$.
\item Calculate $R_{new}=\sum_{s\in\hat{\mathcal{S}}}R_{QD,s}(\hat{\mathcal{S}})$.
\item If $R_{new}>R_{old}$, set $R_{old}=R_{new}$, $\mathcal{S}=\hat{\mathcal{S}}$
\item Set $\hat{u}=\hat{u}+1$ mod $U$.
\end{enumerate}
\item Let $u_h=u_h+|\mathcal{S}|\mbox{ mod }U$.
\end{enumerate}
\end{table}

\end{document}